\documentstyle[aps,prb,epsf,preprint,tighten]{revtex}

\begin{document}

\title{Parity Effect in a Small Superconducting Particle}
\author{S. D. Berger and B. I. Halperin}
\address{Department of Physics, Harvard University, Cambridge, Massachusetts 02138}
\date{\today}
\maketitle

\begin{abstract}
Matveev and Larkin calculated the parity effect on the ground state energy
of a small superconducting particle in the regimes where the mean level spacing 
$\delta $ is either large or small compared to the bulk gap $\Delta .$ We perform a numerical
calculation which extends their results to intermediate values of $\delta
/\Delta .$
\end{abstract}

\section{INTRODUCTION}

\smallskip

Black, Ralph, and Tinkham \cite{BRT,BRT2} performed tunneling experiments on nanometer
scale particles of aluminum at temperatures $T\approx 50$ mK well below the
superconducting transition temperature $T_{c}\sim 1$ K. In some particles 
\cite{BRT} they found a superconducting energy gap larger than the level
spacing. This gap was slightly enhanced from the bulk gap. In smaller
particles \cite{BRT2}, the level spacing exceeded the bulk gap and no
evidence of superconductivity was found. One may then ask whether all
evidence of superconductivity is extinguished in these smaller grains.

Matveev and Larkin \cite{ML} addressed this question within a BCS model 
\begin{equation}
\begin{array}{lll}
H & = & \sum\limits_{j,\sigma }\varepsilon _{j}c_{j\sigma }^{+}c_{j\sigma
}-g\sum\limits_{|\varepsilon _{j}|,|\varepsilon _{j^{\prime }}|<\omega
_{d}}c_{j\uparrow }^{+}c_{j\downarrow }^{+}c_{j^{\prime }\downarrow
}c_{j^{\prime }\uparrow }.
\end{array}
\end{equation}
Letting $E_{g}^{(m)}$ be the ground state
energy for the system with $m$ electrons, they defined the parity effect
parameter 
\begin{equation}
\begin{array}{lll}
\Delta _{P} & = & E_{g}^{(2n+1)}-\frac{1}{2}\left(
E_{g}^{(2n)}+E_{g}^{(2n+2)}\right) .
\end{array}
\label{p}
\end{equation}
This parameter measures the degree to which the odd parity ground states
have a higher energy than the even ones. Matveev and Larkin studied the
limit $g\downarrow 0$, $\omega _{d}\rightarrow \infty $, at fixed mean
level spacing $\delta $
and at fixed value of the bulk gap 
\begin{equation}
\Delta \equiv 2\omega _{d}e^{-\delta /g}.  \label{delta}
\end{equation}
They found the asymptotic results 
\begin{mathletters}
\begin{eqnarray}
\Delta _{P}/\Delta  &\sim &1-\frac{\delta /\Delta }{2},\ \;\quad \delta /%
\Delta <<1\medskip \medskip   \label{stronglimit} \\
\Delta _{P}/\Delta  &\sim &\frac{\delta /\Delta }{2\ln (\delta /\Delta )}%
,\quad \delta /\Delta >>1  \label{weaklimit}
\end{eqnarray}
A minimum was therefore expected in $\Delta _{P}/\Delta $ for $\delta /%
\Delta $ near$\ 1$.

We perform a numerical 
calculation for $\Delta _{P}/\Delta $ with constant level
spacing $\delta$, and locate
the position and value of the minimum. We also verify the
asymptotic limits of $\Delta _{P}/\Delta $ are given by (\ref{stronglimit})
and (\ref{weaklimit}). 

\medskip

\section{METHOD}

\smallskip

We consider a collection of states $j\in J=\{-n,\ldots ,n\}$ about the Fermi
surface and study the effective Hamiltonian for these states 
\end{mathletters}
\begin{equation}
\begin{array}{lll}
\widetilde{H} & = & \delta \sum jc_{j\sigma }^{+}c_{j\sigma }-\widetilde{g}%
\sum c_{j\uparrow }^{+}c_{j\downarrow }^{+}c_{j^{\prime }\downarrow
}c_{j^{\prime }\uparrow }.
\end{array}
\label{H(eff,orig)}
\end{equation}
The effective coupling $\widetilde{g}$ is given in perturbation theory by 
\begin{equation}
\widetilde{g}=\frac{g}{1-\frac{g}{2}\sum\limits_{k\in K}\frac{1}{|k\delta
-\mu |}},  \label{g'}
\end{equation}
where $K$ is the set of integers $k$ satisfying $n<|k|<\omega _{d}/\delta $
and $\mu $ is the chemical potential. We use the effective coupling $%
\widetilde{g}$ in the limit $g\downarrow 0$, $\omega _{d}\rightarrow \infty $
taken at fixed $\Delta $ and fixed $\delta $. The coupling $\widetilde{g}$
can then be written in the form 
\begin{equation}
\widetilde{g}=\frac{\delta }{\ln (a_{n,\mu }\frac{\left( 2n+1\right) \delta 
}{\Delta })}  \label{g'(an)}
\end{equation}
where the constant $a_{n,\mu }\approx 1$ for $n>2$ and $|\mu |<\delta .$ We
nevertheless use the exact $a_{n,\mu }$ in this calculation.

We are interested in the ground state energy of $\widetilde{H}$ for both odd
and even electron number. In the even case the ground state has no
singly-occupied level and can therefore be mapped onto the ground state of
the spin Hamiltonian 
\begin{equation}
\widetilde{H}=\delta \sum j\sigma _{j}^{z}-\widetilde{g}\sum \sigma
_{j}^{+}\sigma _{j^{\prime }}^{-}  \label{H(eff)}
\end{equation}
for $j\in J,$ where the spin operators are defined by 
\begin{equation}
\begin{array}{lll}
\sigma _{j}^{z} & = & c_{j\uparrow }^{+}c_{j\uparrow }+c_{j\downarrow
}^{+}c_{j\downarrow }-1 \\ 
\sigma _{j}^{+} & = & c_{j\uparrow }^{+}c_{j\downarrow }^{+} \\ 
\sigma _{j}^{-} & = & c_{j\downarrow }c_{j\uparrow }
\end{array}
\label{sigdef}
\end{equation}
The odd case is similar except for one singly-occupied level ($j=0$) at the
Fermi surface. This level is inert and can be ignored for our purposes. We
therefore use the effective Hamiltonian (\ref{H(eff)}) with $j\in J-\{0\}$
in the odd case.

We thus define the Hamiltonians $\widetilde{H}^{(2n)},\widetilde{H}^{(2n+1)},
$ and $\widetilde{H}^{(2n+2)}$ by (\ref{H(eff)}) and the following
constraints: 
\begin{equation}
\left\{ 
\begin{array}{lllllllllllll}
\widetilde{H}^{(2n)} & : & \sum \sigma _{j}^{z} & = & -1 & ; & j & \in  & J
& ; & \mu ^{(2n)} & = & -\delta /2 \\ 
\widetilde{H}^{(2n+1)} & : & \sum \sigma _{j}^{z} & = & 0 & ; & j & \in  & 
J-\{0\} & ; & \mu ^{(2n+1)} & = & 0 \\ 
\widetilde{H}^{(2n+2)} & : & \sum \sigma _{j}^{z} & = & 1 & ; & j & \in  & J
& ; & \mu ^{(2n+2)} & = & \delta /2
\end{array}
\right. 
\end{equation}
The couplings entering the $\widetilde{H}^{(m)}$ are $\widetilde{g}_{even}$
for $m=2n,2n+2$ and $\widetilde{g}_{odd}$ for $m=2n+1.$ These are given by
expression (\ref{g'(an)}). As $n$ increases, the ground state energy of $%
\widetilde{H}^{(m)}$ approaches that of the original Hamiltonian $H$ with $m$
electrons, up to a constant which is the same for all $m$.

Since the Hamiltonians $\widetilde{H}^{(2n)}$ and $\widetilde{H}^{(2n+2)}$
are related by particle-hole symmetry $\wp $, which sends $\sigma
^{z}\rightarrow -\sigma ^{z}$ and $\sigma ^{\pm }\rightarrow \sigma ^{\mp },$%
\begin{equation}
\wp \widetilde{H}^{(2n+2)}\wp =\widetilde{H}^{(2n)}-\widetilde{g}_{even},
\end{equation}
we need calculate only the ground state energies $E_{g}^{(2n)}$ and $%
E_{g}^{(2n+1)}.$ We do this by exact diagonalization and form the
combination 
\begin{equation}
\begin{array}{lll}
\Delta _{P} & = & E_{g}^{(2n+1)}-\frac{1}{2}\left(
E_{g}^{(2n)}+E_{g}^{(2n+2)}\right)  \\ 
& = & E_{g}^{(2n+1)}-E_{g}^{(2n)}+\frac{\widetilde{g}_{even}}{2}
\end{array}
\end{equation}

\medskip

\section{RESULTS}

\smallskip

In Figure 1 we plot the results for intermediate values of $\delta /\Delta $%
. The curves from top to bottom are the results for $n=3,4,5,$ and $6,$
respectively, and the straight line is the strong coupling asymptote. We
find excellent convergence in the region near the minimum by $n=6$. For
constant level distribution, we find the 
minimum $\Delta _{P}/\Delta \approx 0.7$ occurs at $\delta /\Delta \approx
0.9$.

In Figure 2 we show the agreement with the weak coupling asymptote, where $%
\Delta _{p}(asymptotic)$ is given by expression (\ref{weaklimit}).

Note. As this manuscript was being completed, there appeared a manuscript by
Mastellone {\em et al.} \cite{MAST} which obtains similar results.

\medskip

\section*{ACKNOWLEDGEMENTS}

\smallskip

We are grateful to D. C. Ralph and M. Tinkham for helpful conversations.
This work was supported by the NSF through Grant No. DMR 94-16910, and
through the Harvard Materials Research Science and Engineering Center, Grant
No. DMR 94-00396.

\begin{center}
\bigskip \rule{4in}{0.02in}
\end{center}

\begin{figure}[tbp]
\epsfxsize=0.9\hsize
\epsffile{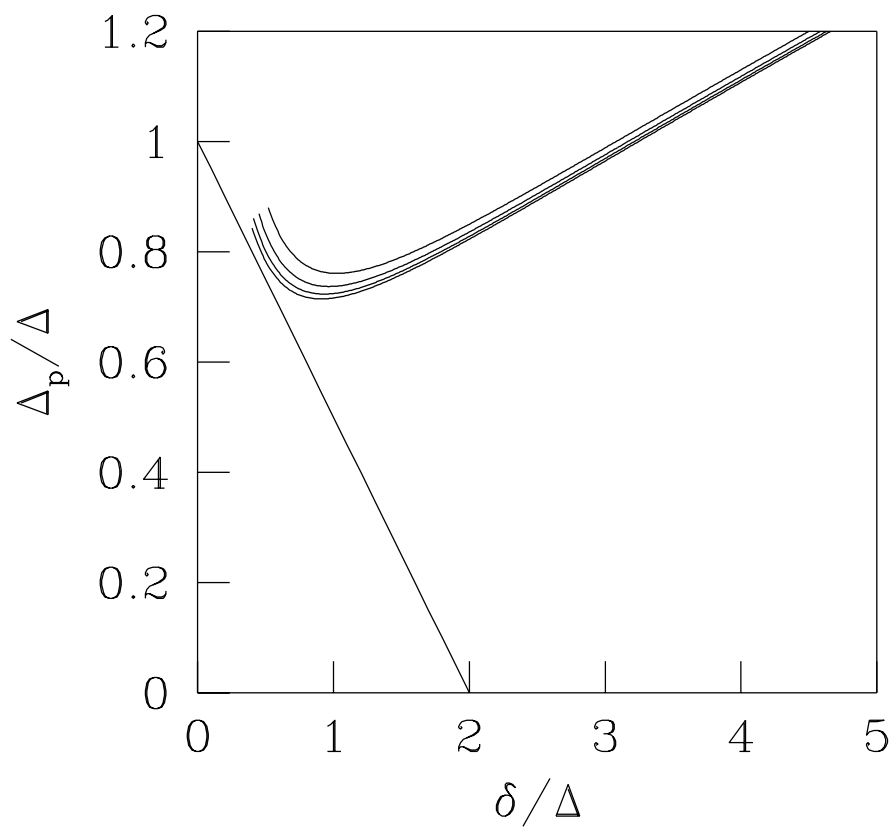}
\caption{The curves from the top to bottom are for $n=3,4,5,$ and $6$. They
approach the strong coupling asymptote indicated by the straight line.}
\label{fig1}%
\end{figure}

\bigskip

\pagebreak 
\begin{figure}[tbp]
\epsfxsize=0.9\hsize
\epsffile{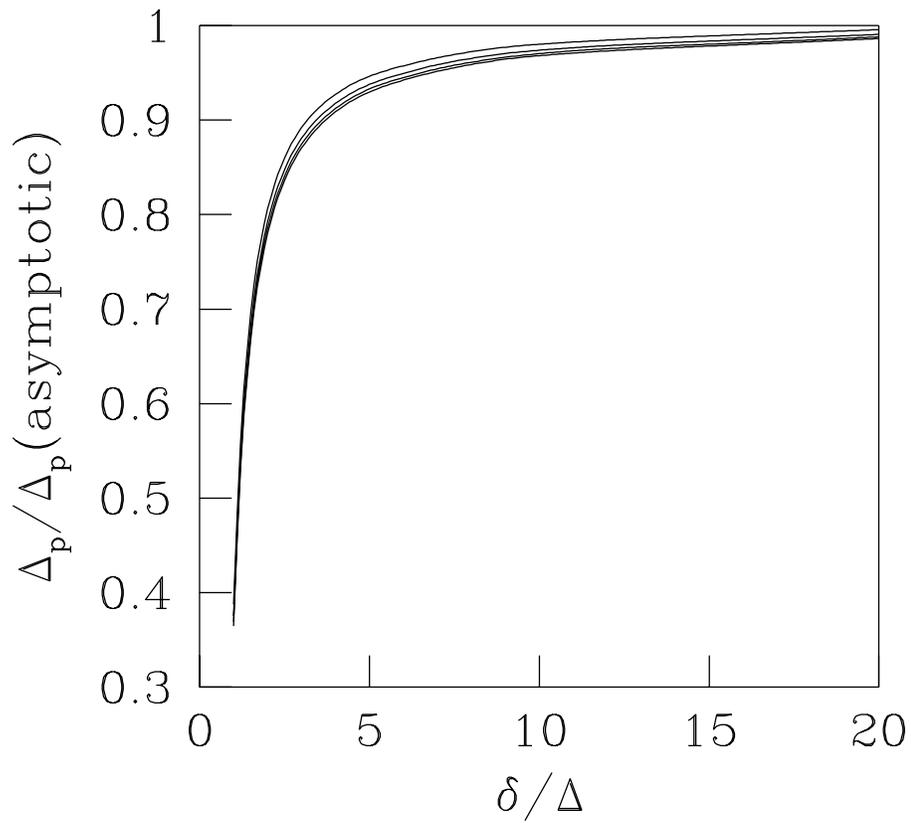}
\caption{The curves from top to bottom are for $n=3,4,5,$ and $6$. All approach
 the weak coupling limit $\Delta_{p}(asymptotic)$ given by expression
(\protect\ref{weaklimit}).}
\label{fig2}%
\end{figure}

\end{document}